\documentclass[english,twocolumn,aps,prb]{revtex4}
\usepackage[T1]{fontenc}
\usepackage[latin9]{inputenc}
\usepackage{graphicx}
\usepackage{amssymb}

\makeatletter

\usepackage{textcomp}

\makeatletter

\makeatletter

\makeatletter


\makeatother

\makeatother

\makeatother

\usepackage{babel}
\makeatother

\begin{document}

\title{Thermally activated exchange narrowing of the Gd$^{3+}$ ESR fine
structure in a single crystal of Ce$_{1-x}$Gd$_{x}$Fe$_{4}$P$_{12}$
($x\approx0.001$) skutterudite.}

\author{\textbf{F. A. Garcia$^{1}$, P. A. Venegas$^{2}$, P. G. Pagliuso$^{1}$,
C. Rettori$^{1,3}$, Z. Fisk$^{4}$, P. Schlottmann$^{5}$, and S.
B. Oseroff$^{6}$}}

\address{$^{1}$Instituto de Física {}``Gleb Wataghin'', C.P. 6165, UNICAMP,
Campinas-SP, 13083-970, Brazil.\\
 $^{2}$UNESP-Universidade Estadual Paulista, Departamento de Física,
Faculdade de Ciências, C.P. 473, 17033-360 Bauru-SP, Brazil\\
 $^{3}$Centro de Ciências Naturais e Humanas, Universidade Federal
do ABC, Santo Andre, SP, 09210-170, Brazil.\\
 $^{4}$University of California, Irvine, CA, 92697-4573, USA.\\
 $^{5}$Department of Physics, Florida State University, Tallahassee,
FL 32306, USA.\\
 $^{6}$San Diego State University, San Diego, California 92182, USA.}

\begin{abstract}
We report electron spin resonance (ESR) measurements in the Gd$^{3+}$
doped semiconducting filled skutterudite compound Ce$_{1-x}$Gd$_{x}$Fe$_{4}$P$_{12}$
($x\approx0.001$). As the temperature $T$ varies from $T\simeq$
150 K to $T\simeq$ 165 K, the Gd$^{3+}$ ESR fine and hyperfine structures
coalesce into a broad inhomogeneous single resonance. At $T\simeq$
200 K the line narrows and as $T$ increases further, the resonance
becomes homogeneous with a thermal broadening of 1.1(2) Oe/K. These
results suggest that the origin of these features may be associated
to a subtle interdependence of thermally activated mechanisms that
combine: \emph{i}) an increase with $T$ of the density of activated
conduction-carriers across the $T$-dependent semiconducting pseudogap;
\emph{ii}) the Gd$^{3+}$ Korringa relaxation process due to an exchange
interaction, $J_{fd}\textbf{S.s}$, between the Gd$^{3+}$ localized
magnetic moments and the thermally activated conduction-carriers and;
\emph{iii}) a relatively weak confining potential of the rare-earth
ions inside the oversized (Fe$_{2}$P$_{3}$)$_{4}$ cage, which allows
the rare-earths to become \emph{rattler} Einstein oscillators above
$T\approx$ 148 K. We argue that the \emph{rattling} of the Gd$^{3+}$
ions, via a motional narrowing mechanism, also contributes to the
coalescence of the ESR fine and hyperfine structure.
\end{abstract}
\maketitle

\section{Introduction}

The filled skutterudite RT$_{4}$X$_{12}$ compounds, where R is a
rare-earth or actinide, T is a transition metal (Fe, Ru, Os) and X
is a pnictogen (P, As, Sb) have attracted great attention due their
broad range of physical properties. In particular, they are of interest
to those investigating basic mechanisms of strongly correlated electronic
systems\citep{Goto,Bauer,Dilley} and also to those seeking for more
efficient thermoelectric materials.\citep{Snyder,Sales}

These compounds crystallize in the LaFe$_{4}$P$_{12}$ structure
with space group $Im3$ and local point symmetry T$_{h}$ for the
R ions. The R ions are guests in the oversized rigid (T$_{2}$X$_{3}$)$_{4}$
cages.\citep{Jeitschko} The dynamics of the guest R ions is believed
to be of great importance in the damping of the thermal conductivity
observed in the filled skutterudite compounds.\citep{Lee,Herman}
Moreover, they may also play an important role in the appearance of
heavy fermion behavior and superconductivity.\citep{Goto,Yanagisawa}

Electron spin resonance (ESR) is a sensitive and powerful microscopic
tool that provides information about crystal field (CF) effects, site
symmetries, valencies of paramagnetic ions, $g$-values, and fine
and hyperfine parameters.\citep{Bleaney} In a recent work our group\citep{Garcia}
found ESR to be a sensitive and useful tool to study the dynamics
of the R ions in this family of filled skutterudites. The weak confining
potential on the R ions at the center of the oversized cage allows
them to easily get \emph{off-center} and experience slightly different
local strength and symmetry of the CF which may lead to: \emph{i})
a distribution of the ESR parameters and, \emph{ii}) a \emph{rattling}
of the R ions that, due to motional narrowing effects,\citep{Anderson}
may cause remarkable changes in the observed ESR spectra. In our previous
ESR experiments on Yb$^{3+}$ in Ce$_{1-x}$Yb$_{x}$Fe$_{4}$P$_{12}$,\citep{Garcia}
these two features were observed and the coexistence of two distinct
Yb$^{3+}$ sites was confirmed.

Ogita \emph{et al},\citep{Ogita} performing Raman scattering experiments
on several metallic skutterudite compounds of the RT$_{4}$X$_{12}$
(T = Fe,Ru,Os; X = P,Sb) series, found \emph{resonant} 2$^{nd}$ order
phonon modes associated with the vibration that change the bond length
of the R-X stretching mode. However, in semiconducting CeFe$_{4}$P$_{12}$
the 2$^{nd}$ order phonon modes were found to be \emph{non-resonant}.
Based on their results Ogita \emph{et al}\citep{Ogita} concluded
that there should be a strong coupling between the R-X stretching
modes and the conduction-electrons (\emph{c-e}). Most reports on the
$T$-dependence of the $dc$-resistivity in CeFe$_{4}$P$_{12}$ present
a semiconductor-like behavior.\citep{Meisner} However, the resistivity
is strongly sample dependent and only in some cases it shows metallic
behavior below $T\approx$ 200 K.\citep{Sato} Nevertheless, for most
of the reported samples the conductivity due to thermally activated
carriers predominates above $T\approx$ 200 K. Thus, for the semiconductor
CeFe$_{4}$P$_{12}$ with gap of $\simeq$ 0.15 eV and estimated Debye
temperature of $\Theta_{D}\simeq$ 500 K,\citep{Aoki} at least a
weak coupling of the R-X stretching mode and the \emph{c-e} should
be expected. This compound experiences a huge increase in the density
of thermally activated conduction-carriers at $T\approx$ 150 K. \citep{Meisner}
Also, evidences for \emph{rattling} of the Yb$^{3+}$ and Ce$^{4+}$
ions were found in ESR\citep{Garcia} and extended x-ray absorption
fine structure (EXAFS)\citep{Cao} experiments, respectively. The
aim of this work is to learn if the presence of thermally activated
conduction-carriers and \emph{rattling} of the R ions can be observed
by the ESR technique. For that reason we measured the evolution of
the Gd$^{3+}$ ESR spectra in Ce$_{1-x}$Gd$_{x}$Fe$_{4}$P$_{12}$
with $T$. To compare our data with a nonrattling compound, we have
also studied the evolution of the Gd$^{3+}$ ESR spectra in Ca$_{1-x}$Gd$_{x}$B$_{6}$
($x\approx0.001$) with $T$, which is a cubic CsCl type semiconductor
with a gap of $\simeq$ 0.8 eV and Debye temperature of $\Theta_{D}\simeq$
783 K.\citep{Vonlanthen}

We found that for Ce$_{1-x}$Gd$_{x}$Fe$_{4}$P$_{12}$ the ESR spectra
show a different behavior in three $T$-regions. At low-$T$ the system
behaves as an insulator, at high-$T$ as a metal and in the intermediate
region it presents the effects of: \emph{a}) an exchange interaction,
$J_{fd}\textbf{S.s}$, between Gd$^{3+}$ localized magnetic moments
and thermally activated conduction-carriers and \emph{b}) possible
evidence for \emph{rattling} of the R ions.

\section{Experimental}

Single crystals of Ce$_{1-x}$Gd$_{x}$Fe$_{4}$P$_{12}$ ($x\lesssim0.001$)
were grown in Sn-flux as described in Ref. \onlinecite{Meisner}.
The cubic structure ($Im3$) and phase purity were checked by x-ray
powder diffraction. Crystals of $\sim$$2$x$2$x$2$ mm$^{3}$ of
naturally grown crystallographic faces were used in the ESR experiments.
Single crystals of Ca$_{1-x}$Gd$_{x}$B$_{6}$ ($x\lesssim0.001$)
were grown as described in Ref. \onlinecite{Young}. The cubic structure
(space group 221, \textit{Pm3m}, CsCl type and local point symmetry
T$_{d}$ for the R ions) and phase purity were checked by x-ray powder
diffraction and the crystals orientation determined by Laue x-ray
diffraction. Most of the ESR experiments were done in $\sim2$x$1$x$0.5$
mm$^{3}$ single crystals. The ESR spectra were taken in a Bruker
X-band ($9.48$ GHz) spectrometer using appropriated resonators coupled
to a $T$-controller of a helium gas flux system for $4.2\lesssim T\lesssim300$
K. The Gd concentrations were determined from the $H$ and $T$-dependence
of the magnetization, $M(H,T)$, measured in a Quantum Design SQUID
$dc$-magnetometer. In both systems the magnetic susceptibility follows
a Curie-Weiss behavior. Also, in both compounds the $T$-dependence
of the Gd$^{3+}$ ESR intensity presents a Curie-Weiss like behavior
within the accuracy of the experiments.

\section{Results and Discussion}

In both compounds at low-$T$ the Gd$^{3+}$ ESR spectra show the
full resolved fine structure corresponding to the spin Hamiltonian
for the Zeeman and cubic CF interactions, ${\cal H}=g\beta HS+b_{4}O_{4}+b_{6}O_{6}$.\citep{Bleaney}
The angular and $T$-dependence of the spectra were taken mostly with
the applied magnetic field, $H$, in the $(1,-1,0)$ plane. The fitting
of the data to the spin Hamiltonian shows that the parameters are,
within the experimental accuracy, $T$-independent for the entire
studied $T$-range. The measured parameters were, $g$ = 1.986(3)
and $b_{4}$ = 7(1) Oe for Ce$_{1-x}$Gd$_{x}$Fe$_{4}$P$_{12}$
and $g$ = 1.992(3) and $b_{4}$ = 13.8(5) Oe for Ca$_{1-x}$Gd$_{x}$B$_{6}$,
in agreement with previous low-$T$ reports.\citep{Mesquita,Urbano}
The accuracy of the data was not enough to estimate the value of $b_{6}$.
The $g$-shift measured for Gd$^{3+}$ in Ce$_{1-x}$Gd$_{x}$Fe$_{4}$P$_{12}$
is negative, $\Delta g$ = 1.986(3) - 1.993 $\approx$ - 0.007. An
additional term to the spin Hamiltonian, $J_{fd}$\textbf{S.s}, due
to a covalent exchange hybridization between the Gd$^{3+}$ 4$f$
electrons and \emph{c-e} with $d$ character would be responsible
for this negative $g$-shift.\citep{Barberis} For the Ce$_{1-x}$Gd$_{x}$Fe$_{4}$P$_{12}$
crystal, careful measurements of the spectra were taken from $T\simeq150$
K to $T\simeq$ 200 K for various directions of $H$. In this $T$-interval,
the fine structure coalesces into a single broad line and its lineshape
changes from lorentzian (insulator) to dysonian (metallic).\citep{Feher}
Notice that these features are independent of the field orientation
and none of them is observed in Ca$_{1-x}$Gd$_{x}$B$_{6}$.

Figures 1, 2 and 3 display the evolution with $T$ (4.2 $\lesssim T\lesssim$
300 K) of the normalized ESR spectra of Gd$^{3+}$ in the Ce$_{1-x}$Gd$_{x}$Fe$_{4}$P$_{12}$
and Ca$_{1-x}$Gd$_{x}$B$_{6}$ ($x\lesssim0.001$) crystals for
$H$ in the $(1,-1,0)$ plane along {[}001], 30$^{\circ}$ from {[}001],
and {[}110], respectively. These data show that for $T\gtrsim$ 150
K the $T$-dependence of the Gd$^{3+}$ fine structure is quite different
in both compounds.

\begin{figure}
\includegraphics[scale=0.6]{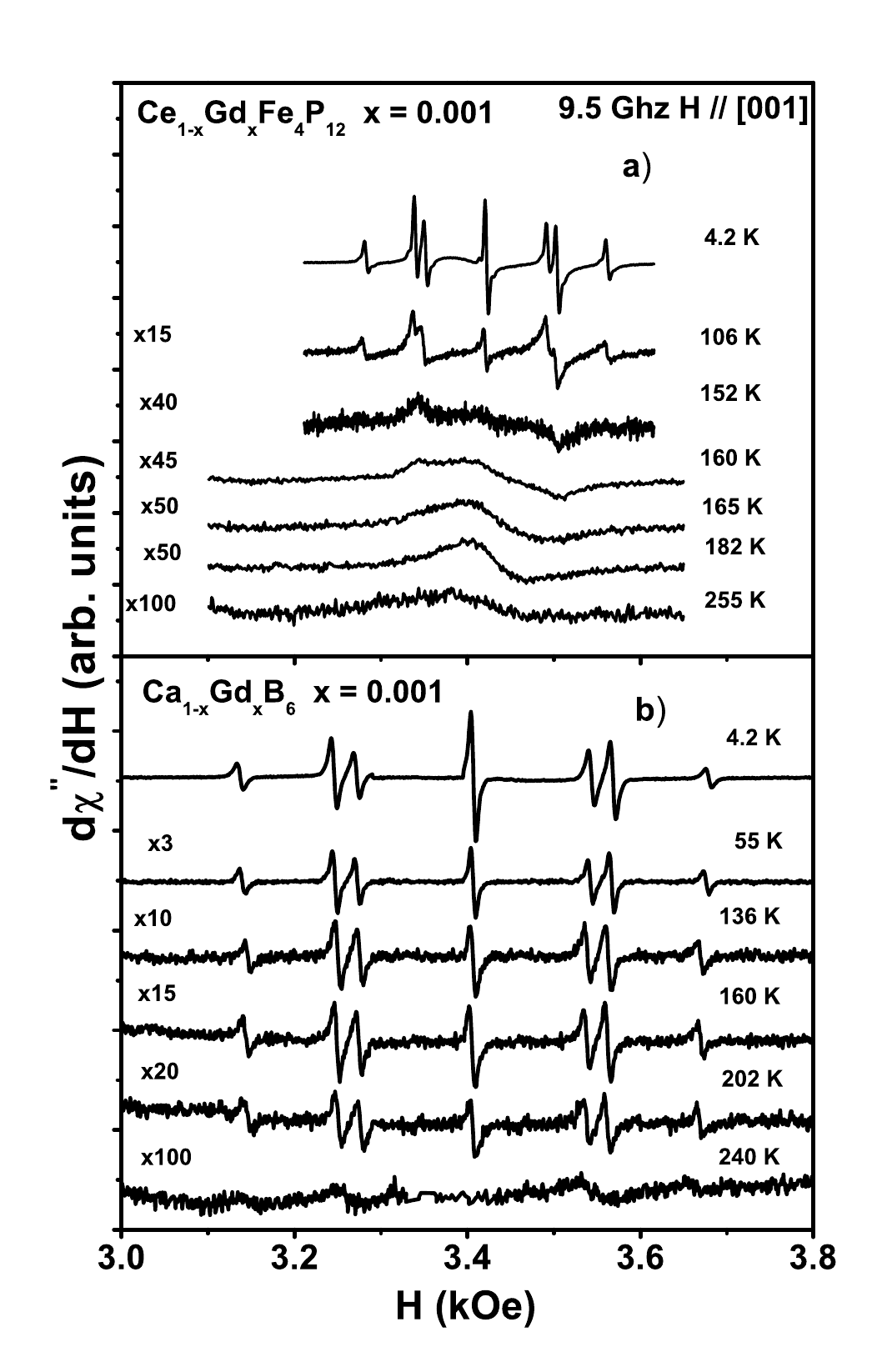}

\caption{$T$-dependence of the X-band ESR spectra for $H\parallel$ {[}001];
a) Ce$_{1-x}$Gd$_{x}$Fe$_{4}$P$_{12}$ and b) Ca$_{1-x}$Gd$_{x}$B$_{6}$.}

\label{fig:1}
\end{figure}

\begin{figure}
\includegraphics[scale=1.6]{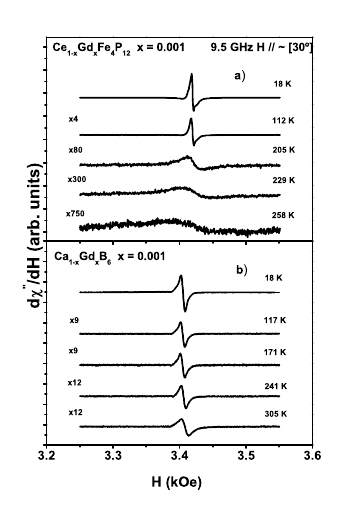}

\caption{$T$-dependence of the X-band ESR spectra for $H$ along $\theta\approx30^{o}$
from {[}001] in the (1,-1,0) plane; a) Ce$_{1-x}$Gd$_{x}$Fe$_{4}$P$_{12}$
and b) Ca$_{1-x}$Gd$_{x}$B$_{6}$.}

\label{fig:2}
\end{figure}

\begin{figure}
\includegraphics[scale=1.6]{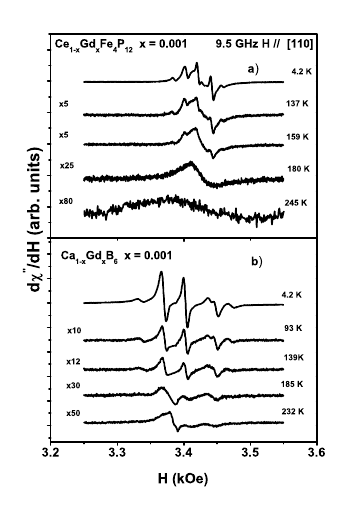}

\caption{$T$-dependence of the X-band ESR spectra for $H\parallel$ {[}110];
a) Ce$_{1-x}$Gd$_{x}$Fe$_{4}$P$_{12}$ and b) Ca$_{1-x}$Gd$_{x}$B$_{6}$.}

\label{fig:3}
\end{figure}

For Ce$_{1-x}$Gd$_{x}$Fe$_{4}$P$_{12}$ the central transition
($\frac{1}{2}\leftrightarrow-\frac{1}{2}$) at 4.2 K and $H$ along
{[}001] is narrow enough to observe the hyperfine satellites lines
of the isotopes $^{155,157}$Gd$^{3+}$ (I = 3/2)(see Fig. 4). The
measured hyperfine parameter is A = 5.5(2) Oe.\citep{FGarcia} This
hyperfine structure is also observed, although not so clearly, for
the other transitions in the spectrum. For the angle where the fine
structure collapses (29.6$^{\circ}$ from {[}001]) and the various
transitions overlap, a small missorientation of $H$ by $\lesssim$
2$^{\circ}$ away from this direction affects the overall lineshape
and the hyperfine structure is then strongly blurred (see Fig. 2).
For Ca$_{1-x}$Gd$_{x}$B$_{6}$ Fig. 4 shows that, due to its higher
$g$-value and broader linewidth, the ($\frac{1}{2}\leftrightarrow-\frac{1}{2}$)
transition is shifted to lower-$H$ and the hyperfine structure is
not well resolved. However, the hyperfine parameter can still be estimated
to be A $\simeq$ 7(1) Oe.

\begin{figure}
\includegraphics[scale=0.5]{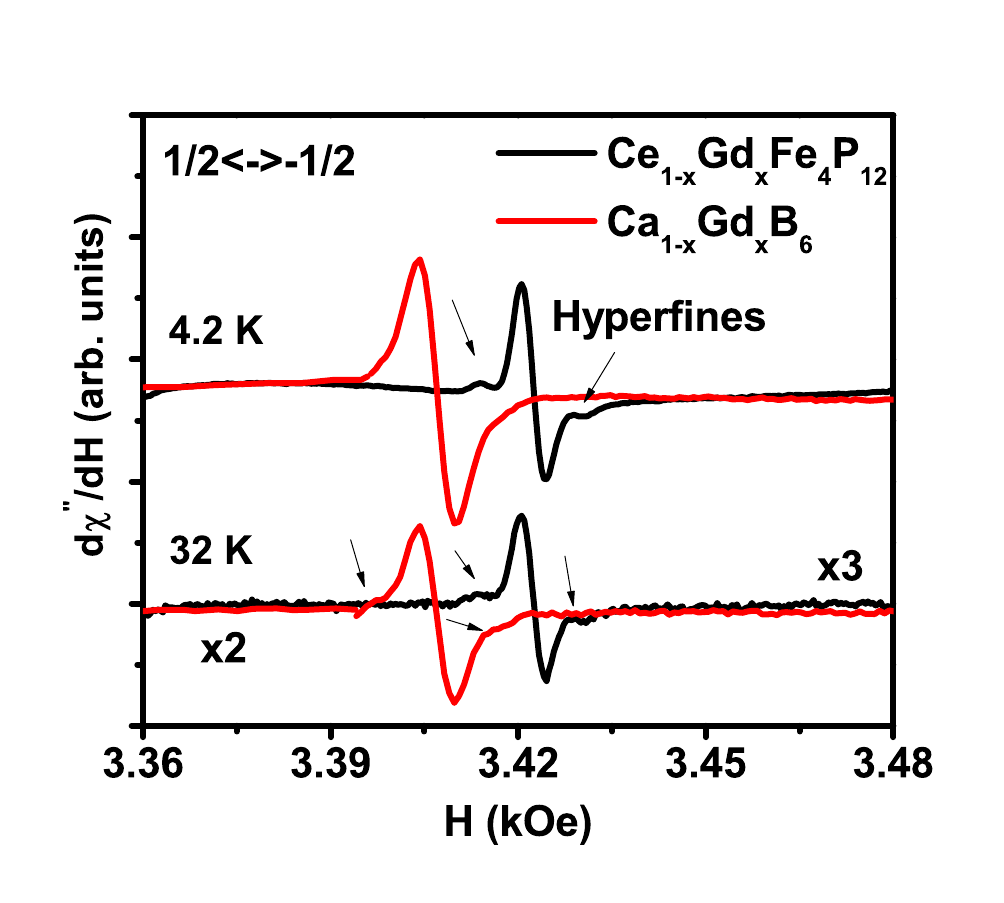}

\caption{X-band ESR ($\frac{1}{2}\leftrightarrow-\frac{1}{2}$) transition
for $H\parallel${[}001]. The arrows show the satellite hyperfine
structure for both samples of $x$ = 0.001.}

\label{fig:4}
\end{figure}

Figure 5 presents for both compounds the evolution with $T$ (4.2
$\lesssim T\lesssim$ 300 K) of the linewidth ($\Delta H$) for the
various Gd$^{3+}$ ESR transitions at several $H$ orientations. For
Ce$_{1-x}$Gd$_{x}$Fe$_{4}$P$_{12}$ Fig. 5a shows the $T$-dependence
of $\Delta H$ for the ($\frac{1}{2}\leftrightarrow-\frac{1}{2}$)
transition and $H$ along {[}001], {[}111], {[}110] directions and
at $\sim$ 30$^{\circ}$ from {[}001] in the $(1,-1,0)$ plane for
the collapsed spectrum. It is clear from the data that there are three
regions of different $T$-dependence of $\Delta H$: Region I, for
$T$ $\lesssim$ 150 K, where $\Delta H$ is nearly $T$-independent
and very narrow at $\simeq$ 5(1) Oe; Region II, for 165 $\lesssim T\lesssim$
200 K, where the full fine structure dramatically coalesces into a
broad inhomogeneous single resonance with anisotropic $\Delta H$;
and Region III, for $T\gtrsim$ 200 K, where $\Delta H$ is again
isotropic and homogeneous. It corresponds to a single coalesced resonance
and has a linear thermal broadening of $\simeq$ 1.1(2) Oe/K, reminiscent
of a Korringa-like relaxation process via the \emph{c-e}.\citep{Korringa}

Figs. 5b and 5c show the $T$-dependence of $\Delta H$ for the various
transitions in Ce$_{1-x}$Gd$_{x}$Fe$_{4}$P$_{12}$ and Ca$_{1-x}$Gd$_{x}$B$_{6}$
for $H$ along {[}001] and 30$^{\circ}$ from {[}001], respectively.
A timid broadening starts to be observed on $\Delta H$ for the fine
structure components at $T\simeq$ 60 K for Ce$_{1-x}$Gd$_{x}$Fe$_{4}$P$_{12}$
and at $T\simeq$ 120 K for Ca$_{1-x}$Gd$_{x}$B$_{6}$. Presumably
this broadening is caused by a phonon spin-lattice relaxation process.\citep{Orbach}
The fact that such a phonon contribution starts at lower-$T$ in Ce$_{1-x}$Gd$_{x}$Fe$_{4}$P$_{12}$
than in Ca$_{1-x}$Gd$_{x}$B$_{6}$ is consistent with the lower
Debye temperature for the former compound. Alternatively, the Gd$^{3+}$
ions produce bound states in the gap, which in the case of CaB$_{6}$
are donor states. Carriers bounded at low-$T$ in these states can
be promoted into the conduction band as $T$ increases and produce
a faster relaxation. However, as $T$ increases in Region I a small
local distribution of CF cannot be excluded as the reason for the
small broadening of the fine structure lines. The large voided space
and concomitant increase of the carrier density as $T$ increases
may thermally activate slow motions of the Gd$^{3+}$ ions inside
the oversized (Fe$_{2}$P$_{3}$)$_{4}$ cage which could slightly
alter, in an inhomogeneous way, the local CF at the Gd$^{3+}$ site.

\begin{figure}
\includegraphics[scale=0.75]{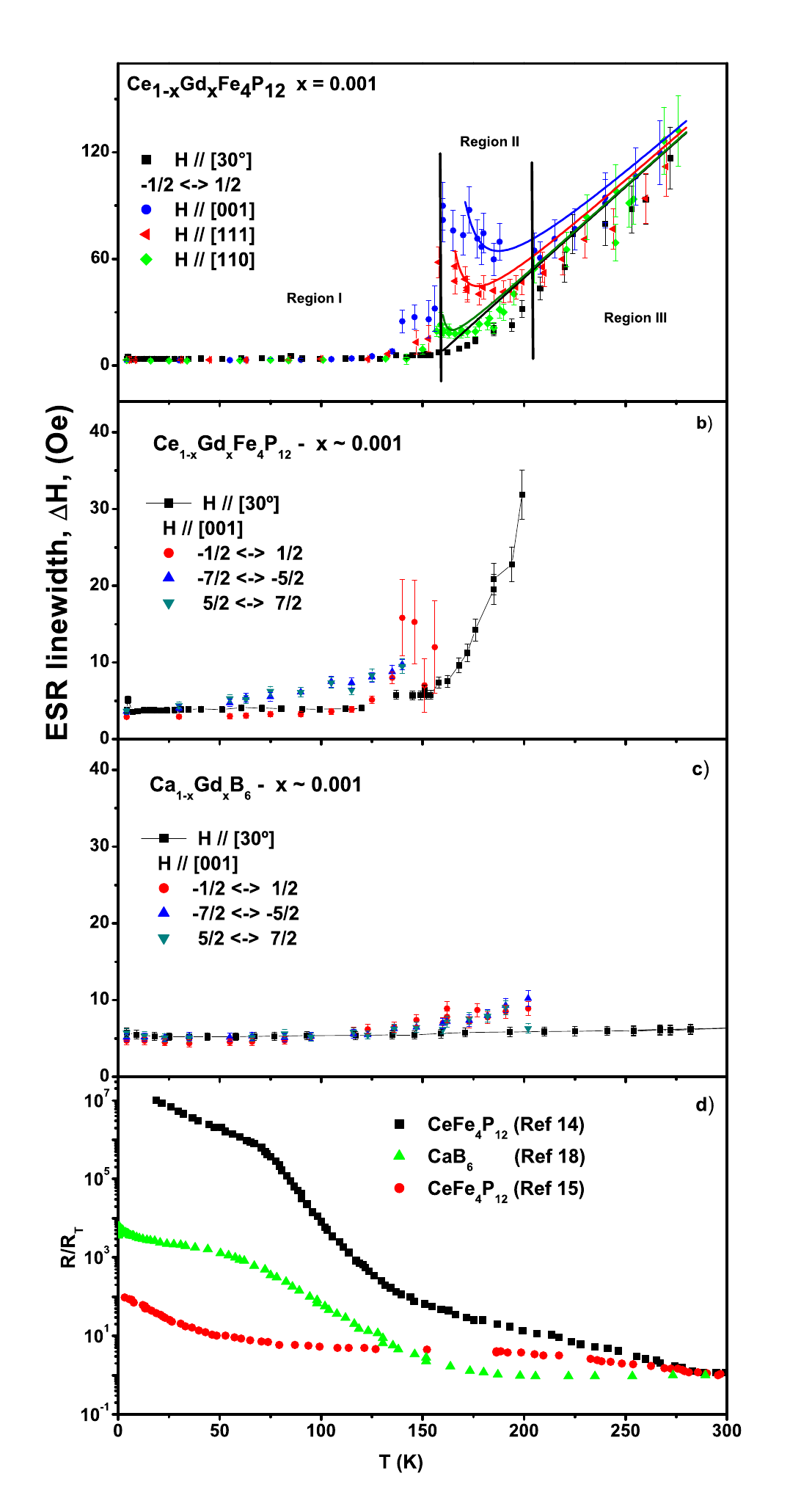}

\caption{$T$-evolution (4.2 $\lesssim T\lesssim$ 300 K) of the Gd$^{3+}$
ESR linewidth, $\Delta H$, for both compounds and various transitions
at several $H$ orientations. Notice that in Fig. 5a the behavior
of $\Delta H(T)$ clearly characterizes three different Regions (I,
II, and III). The solid lines in Regions II and III correspond the
calculated $\Delta H(T)$ for the coalesced ESR spectra using the
Plefka-Barnes\citep{Plefka} exchange narrowing mechanism (see text).
Fig. 5d presents the general $T$-dependence reported for the resistivity
in these compounds. The resistivity of our crystals is similar to
that of Ref. 14.}

\label{fig:5}
\end{figure}

As already mentioned, for Ce$_{1-x}$Gd$_{x}$Fe$_{4}$P$_{12}$ above
$T\simeq$ 160 K a dramatic broadening mechanism drives the whole
Gd$^{3+}$ resolved ESR fine structure in Region I to coalesce into
the broad inhomogeneous and unresolved anisotropic spectrum of Region
II (see Fig. 5a). This striking result occurs at about the same $T$
where: \emph{i)} the density of thermally activated mobile carriers
increases by several orders of magnitude (see Fig. 5d); \emph{ii)}
the rattling of the filler R atom is confirmed by EXAFS experiments;\citep{Cao}
\emph{iii)} the existence of a $T$-dependent semiconducting pseudogap
is observed for $T\lesssim$ 300 K in ultraviolet and x-ray photoemission
spectroscopies (UPS, XPS),\citep{Rayjada} and; \emph{iv)} where the
change from lorentzian (insulator) to dysonian (metallic) ESR lineshape
is observed (see Figs. 1a, 2a and 6). Note that none of these features
are present in Ca$_{1-x}$Gd$_{x}$B$_{6}$. The solid lines in Regions
II and III of Fig. 5a are the calculated $\Delta H(T)$ for the coalescing
ESR spectra using the Plefka-Barnes\citep{Plefka} exchange ($J_{fd}$\textbf{S.s})
narrowing theory of the fine structure. In the calculation we used
a Korringa relaxation of 1.1(2) Oe/K that is \emph{{}``switched-on''}
at 157(2) K, a fourth order CF parameter $b_{4}$ = 7(1) Oe and a
residual linewidth $\Delta H(T=0)$ = 5(1) Oe.

Figure 6 presents the $T$-dependence of the hyperfine structure for
the ($\frac{1}{2}\leftrightarrow-\frac{1}{2}$) transition. The data
show that the coalescence of the hyperfine structure is already observed
at $T\approx$ 150 K, i.e., at $\approx$ 15 K below the coalescence
of the fine structure at $T\approx$ 165 K. This is expected since
the exchange interaction would act first on the hyperfine structure
due to its much smaller spectral splitting.

\begin{figure}
\includegraphics[scale=0.6]{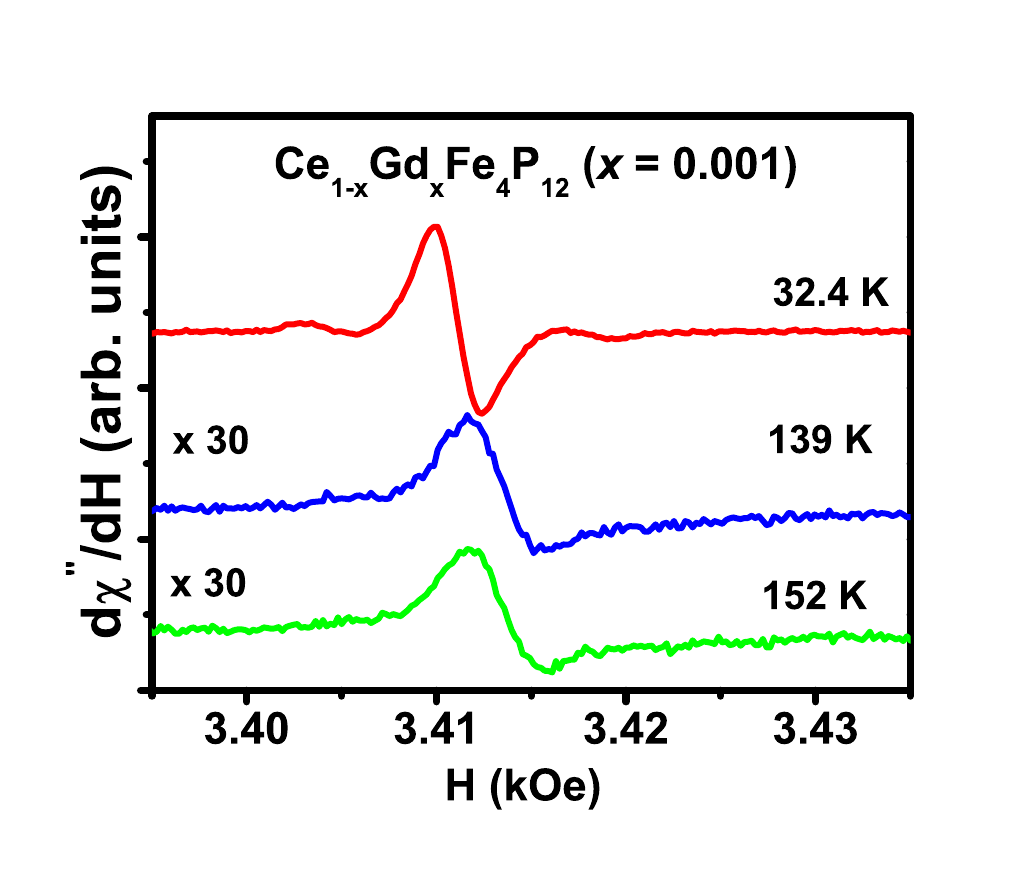}

\caption{$T$-dependence of the hyperfine structure for the ($\frac{1}{2}\leftrightarrow-\frac{1}{2}$)
transition. Notice that the apparent difference in the field for resonance
between the low and high-$T$ spectra is mainly due to the change,
from lorenztian to dysonian, and also to a small change in the frequency
of the microwave cavity due to the temperature ($\lesssim$0.05$\%$).}

\label{fig:6} 
\end{figure}

Figure 7 shows the angular dependence of $\Delta H$ at different
$T$ corresponding to Regions II and III. Following the analysis of
Urban \textit{et al.}\citep{Urban} for the exchange narrowing of
the Gd$^{3+}$ ESR fine structure, the anisotropy of $\Delta H$ in
Region II can be fitted to the general expression for the intermediate
coupling regime:

\begin{equation}
\Delta H=A(T)+B(T)p^{2}(\theta),\label{H}\end{equation}
 where \begin{equation}
p^{2}(\theta)=1-5[\sin^{2}(\theta)+(3/4)\sin^{4}(\theta)].\label{P}\end{equation}

Fig. 5a shows that there is narrowing of $\Delta H$ for $T$ approaching
200 K and that the anisotropy decreases, \emph{i.e}., $B\rightarrow$
0 as $T\rightarrow200$. For $T\gtrsim$ 200 K $\Delta H$ becomes
isotropic (see Fig. 7) and increases linearly at a rate of 1.1(2)
Oe/K (see Fig. 5a). This linear increase is an evidence for the presence
of a Korringa relaxation process, i.e., the Gd$^{3+}$ ions relax
to the lattice via an exchange interaction, $J_{fd}\textbf{S.s}$,
between the Gd$^{3+}$ localized magnetic moment and the thermally
activated conduction-carriers.\citep{Korringa}

A $g$-shift of $\Delta g=-0.007$ has been measured for the entire
range of $T$ studied for Ce$_{1-x}$Gd$_{x}$Fe$_{4}$P$_{12}$.
This is surprising because in Region I there are no conduction-carriers
that could be polarized. However, the host is a Kondo insulator with
a finite Van Vleck susceptibility due to the crystalline field splitting
of the Ce ions. This Van Vleck susceptibility is larger than the susceptibility
of the thermally excited electrons in Region II and III and provides
the polarization to produce the $g$-shift. This effect is of course
not present in the Ca$_{1-x}$Gd$_{x}$B$_{6}$ sample, since CaB$_{6}$
has no significant susceptibility. There is a second unusual issue
with the $g$-shift. In a simple metallic host Gd$^{3+}$ ions are
expected to have a ferromagnetic Heisenberg exchange. However, the
$g$-shift is negative, indicative of a hybridization mechanism. The
overlap of the Gd $4f$ electrons with the hybridized Ce $4f$-band
forming the valence and conduction bands of the Kondo insulator could
give rise to an antiferromagnetic exchange.

It is interesting to note that in Gd$^{3+}$ doped simple metals the
Korringa relaxation, d($\Delta H$)/d$T$, would be related to the
$g$-shift, $\Delta g$, by:\citep{Rettori} \begin{equation}
d(\Delta H)/dT=(\pi k/g\mu_{B})(\Delta g)^{2}.\label{DH}\end{equation}
 Using our experimental value of 1.1(2) Oe/K for d($\Delta H$)/d$T$
and 2.34 x 10$^{4}$ Oe/K for $\pi k$/$g\mu_{B}$ we estimate a corresponding
$\Delta g$ of $\approx$ $\mid$0.007$\mid$. These results and Eq.
(\ref{DH}) suggest that: \emph{i)} in Region I, where there is no
Korringa relaxation, the exchange coupling due to covalent hybridization
gives rise to just polarization effects, $J_{fd}(\textbf{q}=0)$;\citep{Davidov}
\emph{ii)} the trigger of the Korringa mechanism in Regions II and
III is due to the presence of mobile activated conduction-carriers
at the Fermi level, which are responsible for the momentum transfer
between the conduction-carriers and the localized magnetic moment
via the exchange coupling, $J_{fd}(\textbf{q}\neq0)$;\citep{Rettori,Davidov}
and \emph{iii)} in the metallic Regions II and III, there is no $\textbf{q}$-dependence
of the exchange interaction, i.e., $J_{fd}(\textbf{q}=0)\equiv\langle J_{fd}^{2}(\textbf{q})\rangle_{E_{F}}^{1/2}$.\citep{Rettori,Davidov}

\begin{figure}
\includegraphics[scale=0.6]{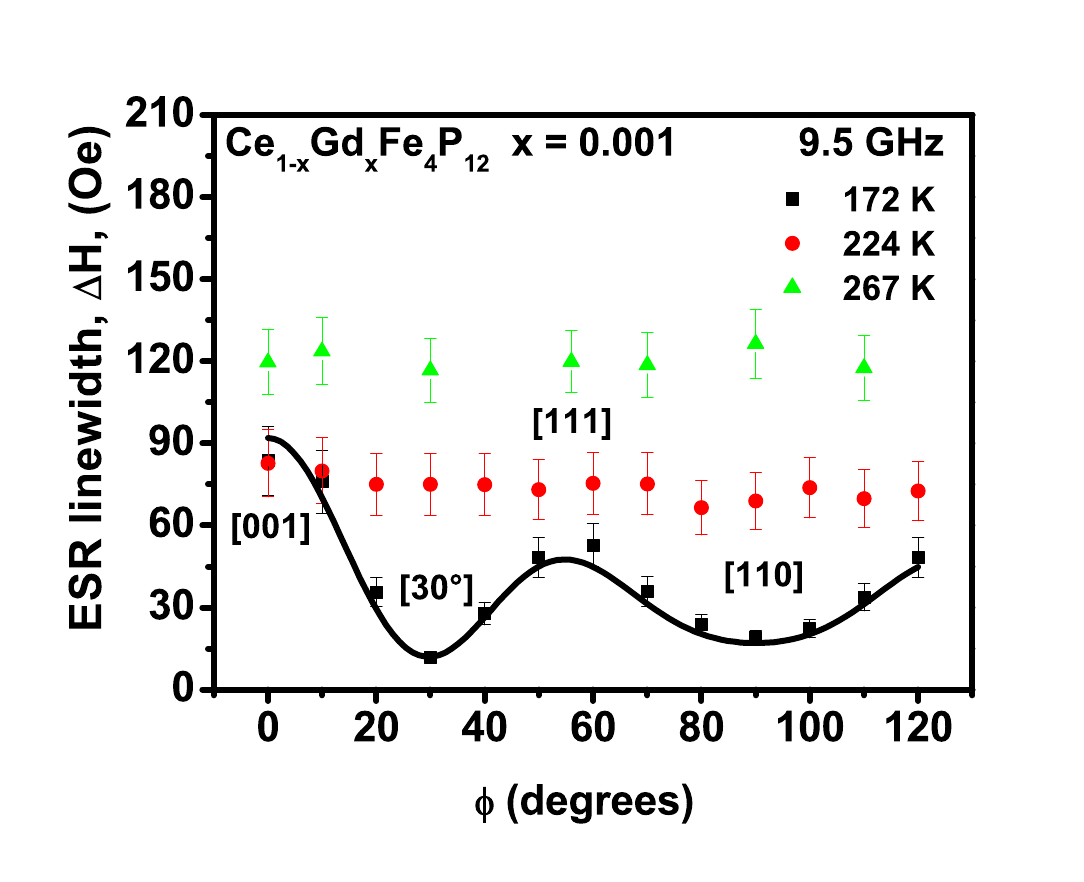}

\caption{Angular dependence of the $\Delta H$ for the Gd$^{3+}$ ESR for temperatures
in the Regions II and III of Fig. 5a. The solid line corresponds to
the fitting of the data to Eq. 1 for A(172 K) = 10(3) Oe and B(172
K) = 80(10) Oe.}

\label{fig:7}
\end{figure}

For $T\lesssim$ 10 K Figure 8 shows that, due to the long spin-lattice
relaxation time $T_{1}$ in these materials, the Gd$^{3+}$ ($\frac{1}{2}\leftrightarrow-\frac{1}{2}$)
transition saturates as a function of the microwave power.\citep{Poole}
From $\Delta H$ = ($\gamma$$T_{2}$)$^{-1}$ the spin-spin relaxation
time can be estimated to be $T_{2}\simeq0.1\mu$s in both compounds.
From the saturation factor, S = {[}1+(1/4)$H_{1}^{2}$$\gamma^{2}$$T_{1}$$T_{2}$]$^{-1}$
and P = (1/4)$H_{1}^{2}$, we estimate the spin-lattice relaxation
time to be $T_{1}\simeq$ 10 ms and $T_{1}\simeq$ 4 ms for Ce$_{1-x}$Gd$_{x}$Fe$_{4}$P$_{12}$
and Ca$_{1-x}$Gd$_{x}$B$_{6}$, respectively. Notice that a Korringa
relaxation is absent in Region I where the compound behaves as an
insulator. Moreover, from $\Delta H$ at $T\simeq$ 300 K in Fig.
5a we estimate $T_{1}\simeq0.002\mu$s which is much shorter than
the low-$T$ value of $T_{2}$. Therefore, at high-$T$ $T_{2}\simeq T_{1}$
and, as far as ESR is concerned, this is another evidence that Ce$_{1-x}$Gd$_{x}$Fe$_{4}$P$_{12}$
behaves as a regular metal in Regions II and III.

\begin{figure}
\includegraphics[scale=0.6]{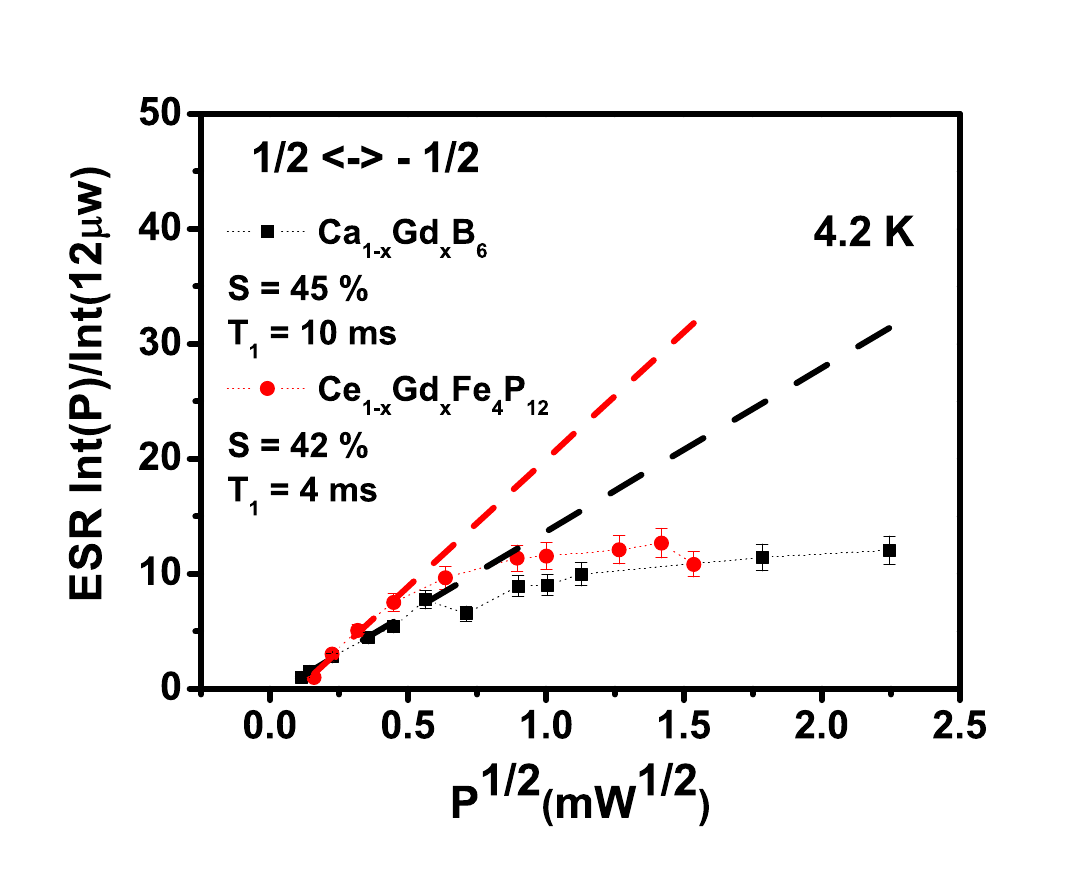}

\caption{Dependence of the Gd$^{3+}$ ESR intensity at low-$T$ on the microwave
power for both materials.}

\label{fig:8}
\end{figure}

Finally, according to the Raman results and conclusions about the
\emph{rattling} modes in metallic skutterudites,\citep{Ogita} it
is plausible that the huge increase in the metallic character of CeFe$_{4}$P$_{12}$
activates the R-X stretching mode and triggers the \emph{rattling}
of the R ions inside the oversized (Fe$_{2}$P$_{3}$)$_{4}$ cage.
Hence, via a motional narrowing mechanism\citep{Anderson} the \emph{rattling}
of the Gd$^{3+}$ ions could also contribute to the dramatic change
of the ESR spectra observed at the transition from Region I to Region
II. Actually Figure 9 shows that the exchange narrowing mechanism
alone cannot reproduce the observed experimental single coalesced
resonance at the transition between these two Regions. Thus, a motional
narrowing of the Gd$^{3+}$ fine structure, due to \emph{rattling}
of the R ions, cannot be disregarded and it probably contributes to
the experimental observed spectra. It should be mentioned that such
a striking behavior is not expected in Ca$_{1-x}$Gd$_{x}$B$_{6}$
due to the much larger semiconducting gap of $\approx$ 0.8 eV and
the tighter cages for the CaB$_{6}$ compound.

\begin{figure}
\includegraphics[scale=0.25]{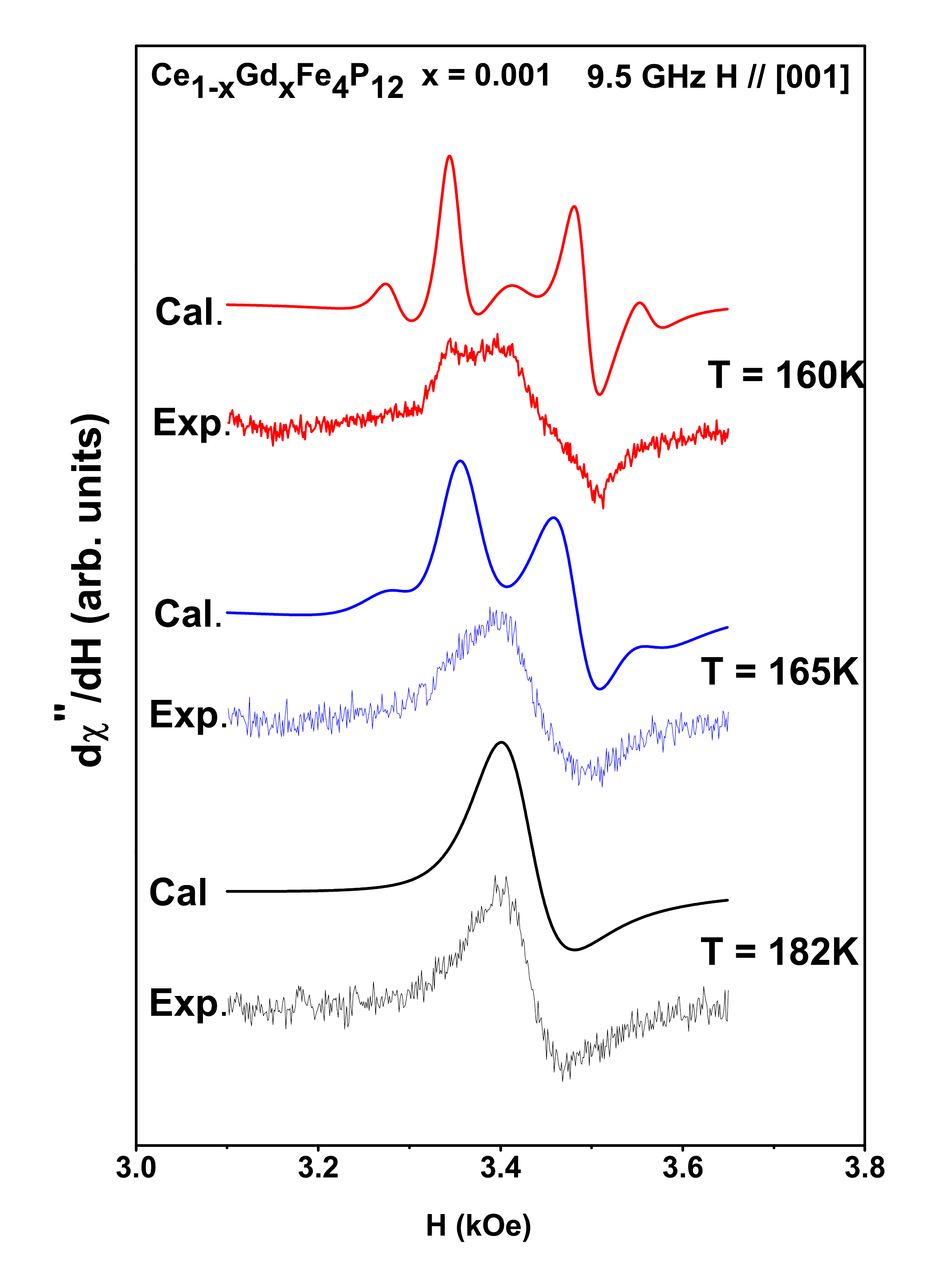}

\caption{Experimental and calculated\citep{Plefka} Gd$^{3+}$ ESR spectra
in Ce$_{1-x}$Gd$_{x}$Fe$_{4}$P$_{12}$ for $x$ = 0.001 at the
transition from Region I to Region II. The parameters used to calculate
the theoretical spectra were the same as those used for the calculation
of $\Delta H(T)$ in Fig. 5a.}

\label{fig:9}
\end{figure}

\section{Conclusions}

In this work we have presented experimental data that show the following
striking features for the $T$-dependence of the Gd$^{3+}$ ESR spectra
in Ce$_{1-x}$Gd$_{x}$Fe$_{4}$P$_{12}$: \emph{a)} the coalescence
of the hyperfine and fine structures at $T\simeq$ 150 K and $T\simeq$
165 K, respectively; \emph{b)} at about these temperatures the ESR
lineshape changes from lorentzian (insulating media) to dysonian (metallic
media); and \emph{c)} the $T$-dependence of the ESR $\Delta H$ changes
from a narrow nearly $T$-independent linewidth for each fine structure
in Region I to a single inhomogeneous broad resonance with anisotropic
$\Delta H$ in Region II and then, in Region III, to a homogeneous
linewidth with a broadening which is linear in $T$, resembling the
Korringa-like relaxation process in a metallic host.\citep{Korringa}
Point \emph{b)} indicates that, at our microwave frequency and between
$T\simeq$ 150 K and $T\simeq$ 165 K, there is also a clear and strong
change in the $ac$-conductivity of the material. We associate this
change to a \emph{smooth crossover} from insulator to metal which
was only possible to be detected due to the high sensitivity that
the ESR lineshape has in a metallic media.

Our ESR observations in Ce$_{1-x}$Gd$_{x}$Fe$_{4}$P$_{12}$, along
with those of Raman, EXAFS, UPS and XPS for CeFe$_{4}$P$_{12}$,
suggest that this \emph{smooth insulator-metal} crossover may be responsible
for the coalescence and narrowing of the hyperfine and fine structures
and also for the activation of the R-X stretching mode that probably
triggers the \emph{rattling} of the Gd$^{3+}$ ions in the oversized
(Fe$_{2}$P$_{3}$)$_{4}$ cage. Via a motional narrowing mechanism
the Gd$^{3+}$ \emph{rattling} may also contribute to the dramatic
change of the ESR spectra at the transition from Region I to Region
II.\citep{Garcia,Anderson}

We believe that our ESR study gives further clues and insights for
the subtle interplay between the local vibration modes (Einstein oscillators)
of the R ions and the \emph{c-e} in the filled skutterudite compounds.
In particular, our work supports the idea that some metallic character
is always needed to set up the necessary conditions for the \emph{rattling}
of the R ions in these materials.

\section{Aknowledgment}

This work was supported in part by FAPESP, CNPq, CAPES and NCC from
Brazil. PS is supported by the US Department of Energy through grant
No. DE-FG02-98ER45707.

\end{document}